\documentclass[pra,amsfonts, twocolumn]{revtex4}

\bibpunct{[}{]}{,}{a}{}{;}
\RequirePackage{times}
\RequirePackage{mathptm}

\begin{document}
\title{Reversible computation as a model for the quantum measurement process\footnote{published in ``Cybernetics and Systems '98 Volume I, ed. by R. Trappl (Austrian Society for Cybernetic Studies, Vienna, 1998), pp. 102-106''}}
\author{Karl Svozil}
\email{svozil@tuwien.ac.at}
\homepage{http://tph.tuwien.ac.at/~svozil}
\affiliation{Institute for Theoretical Physics, Vienna University of Technology,  \\
Wiedner Hauptstra\ss e 8-10/136, A-1040 Vienna, Austria}

\begin{abstract}
One-to-one reversible automata are introduced. Their applicability to a modelling of the quantum mechanical measurement process is discussed.
\end{abstract}

\maketitle

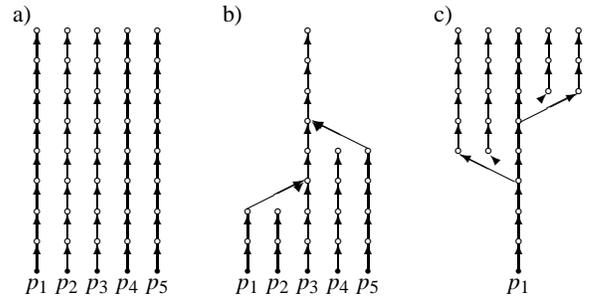
\begin{figure}
\begin{center}
\unitlength 0.40mm
\linethickness{0.4pt}
\begin{picture}(191.00,95.00)
\put(10.00,10.00){\circle*{2.00}}
\put(20.00,10.00){\circle*{2.00}}
\put(30.00,10.00){\circle*{2.00}}
\put(40.00,10.00){\circle*{2.00}}
\put(10.00,20.00){\circle{2.00}}
\put(10.00,11.00){\vector(0,1){8.00}}
\put(10.00,30.00){\circle{2.00}}
\put(10.00,21.00){\vector(0,1){8.00}}
\put(10.00,40.00){\circle{2.00}}
\put(10.00,31.00){\vector(0,1){8.00}}
\put(10.00,50.00){\circle{2.00}}
\put(10.00,41.00){\vector(0,1){8.00}}
\put(10.00,60.00){\circle{2.00}}
\put(10.00,51.00){\vector(0,1){8.00}}
\put(10.00,70.00){\circle{2.00}}
\put(10.00,61.00){\vector(0,1){8.00}}
\put(10.00,80.00){\circle{2.00}}
\put(10.00,71.00){\vector(0,1){8.00}}
\put(10.00,90.00){\circle{2.00}}
\put(10.00,81.00){\vector(0,1){8.00}}
\put(20.00,20.00){\circle{2.00}}
\put(20.00,11.00){\vector(0,1){8.00}}
\put(20.00,30.00){\circle{2.00}}
\put(20.00,21.00){\vector(0,1){8.00}}
\put(20.00,40.00){\circle{2.00}}
\put(20.00,31.00){\vector(0,1){8.00}}
\put(20.00,50.00){\circle{2.00}}
\put(20.00,41.00){\vector(0,1){8.00}}
\put(20.00,60.00){\circle{2.00}}
\put(20.00,51.00){\vector(0,1){8.00}}
\put(20.00,70.00){\circle{2.00}}
\put(20.00,61.00){\vector(0,1){8.00}}
\put(20.00,80.00){\circle{2.00}}
\put(20.00,71.00){\vector(0,1){8.00}}
\put(20.00,90.00){\circle{2.00}}
\put(20.00,81.00){\vector(0,1){8.00}}
\put(30.00,20.00){\circle{2.00}}
\put(30.00,11.00){\vector(0,1){8.00}}
\put(30.00,30.00){\circle{2.00}}
\put(30.00,21.00){\vector(0,1){8.00}}
\put(30.00,40.00){\circle{2.00}}
\put(30.00,31.00){\vector(0,1){8.00}}
\put(30.00,50.00){\circle{2.00}}
\put(30.00,41.00){\vector(0,1){8.00}}
\put(30.00,60.00){\circle{2.00}}
\put(30.00,51.00){\vector(0,1){8.00}}
\put(30.00,70.00){\circle{2.00}}
\put(30.00,61.00){\vector(0,1){8.00}}
\put(30.00,80.00){\circle{2.00}}
\put(30.00,71.00){\vector(0,1){8.00}}
\put(30.00,90.00){\circle{2.00}}
\put(30.00,81.00){\vector(0,1){8.00}}
\put(40.00,20.00){\circle{2.00}}
\put(40.00,11.00){\vector(0,1){8.00}}
\put(40.00,30.00){\circle{2.00}}
\put(40.00,21.00){\vector(0,1){8.00}}
\put(40.00,40.00){\circle{2.00}}
\put(40.00,31.00){\vector(0,1){8.00}}
\put(40.00,50.00){\circle{2.00}}
\put(40.00,41.00){\vector(0,1){8.00}}
\put(40.00,60.00){\circle{2.00}}
\put(40.00,51.00){\vector(0,1){8.00}}
\put(40.00,70.00){\circle{2.00}}
\put(40.00,61.00){\vector(0,1){8.00}}
\put(40.00,80.00){\circle{2.00}}
\put(40.00,71.00){\vector(0,1){8.00}}
\put(40.00,90.00){\circle{2.00}}
\put(40.00,81.00){\vector(0,1){8.00}}
\put(50.00,10.00){\circle*{2.00}}
\put(50.00,20.00){\circle{2.00}}
\put(50.00,11.00){\vector(0,1){8.00}}
\put(50.00,30.00){\circle{2.00}}
\put(50.00,21.00){\vector(0,1){8.00}}
\put(50.00,40.00){\circle{2.00}}
\put(50.00,31.00){\vector(0,1){8.00}}
\put(50.00,50.00){\circle{2.00}}
\put(50.00,41.00){\vector(0,1){8.00}}
\put(50.00,60.00){\circle{2.00}}
\put(50.00,51.00){\vector(0,1){8.00}}
\put(50.00,70.00){\circle{2.00}}
\put(50.00,61.00){\vector(0,1){8.00}}
\put(50.00,80.00){\circle{2.00}}
\put(50.00,71.00){\vector(0,1){8.00}}
\put(50.00,90.00){\circle{2.00}}
\put(50.00,81.00){\vector(0,1){8.00}}
\put(80.00,10.00){\circle*{2.00}}
\put(90.00,10.00){\circle*{2.00}}
\put(100.00,10.00){\circle*{2.00}}
\put(110.00,10.00){\circle*{2.00}}
\put(80.00,20.00){\circle{2.00}}
\put(80.00,11.00){\vector(0,1){8.00}}
\put(80.00,30.00){\circle{2.00}}
\put(80.00,21.00){\vector(0,1){8.00}}
\put(90.00,20.00){\circle{2.00}}
\put(90.00,11.00){\vector(0,1){8.00}}
\put(90.00,30.00){\circle{2.00}}
\put(90.00,21.00){\vector(0,1){8.00}}
\put(100.00,20.00){\circle{2.00}}
\put(100.00,11.00){\vector(0,1){8.00}}
\put(100.00,30.00){\circle{2.00}}
\put(100.00,21.00){\vector(0,1){8.00}}
\put(100.00,40.00){\circle{2.00}}
\put(100.00,31.00){\vector(0,1){8.00}}
\put(100.00,50.00){\circle{2.00}}
\put(100.00,41.00){\vector(0,1){8.00}}
\put(100.00,60.00){\circle{2.00}}
\put(100.00,51.00){\vector(0,1){8.00}}
\put(100.00,70.00){\circle{2.00}}
\put(100.00,61.00){\vector(0,1){8.00}}
\put(100.00,80.00){\circle{2.00}}
\put(100.00,71.00){\vector(0,1){8.00}}
\put(100.00,90.00){\circle{2.00}}
\put(100.00,81.00){\vector(0,1){8.00}}
\put(110.00,20.00){\circle{2.00}}
\put(110.00,11.00){\vector(0,1){8.00}}
\put(110.00,30.00){\circle{2.00}}
\put(110.00,21.00){\vector(0,1){8.00}}
\put(110.00,40.00){\circle{2.00}}
\put(110.00,31.00){\vector(0,1){8.00}}
\put(110.00,50.00){\circle{2.00}}
\put(110.00,41.00){\vector(0,1){8.00}}
\put(120.00,10.00){\circle*{2.00}}
\put(120.00,20.00){\circle{2.00}}
\put(120.00,11.00){\vector(0,1){8.00}}
\put(120.00,30.00){\circle{2.00}}
\put(120.00,21.00){\vector(0,1){8.00}}
\put(120.00,40.00){\circle{2.00}}
\put(120.00,31.00){\vector(0,1){8.00}}
\put(120.00,50.00){\circle{2.00}}
\put(120.00,41.00){\vector(0,1){8.00}}
\put(80.00,31.00){\vector(2,1){18.67}}
\put(90.00,31.00){\vector(1,1){8.67}}
\put(110.00,51.00){\vector(-1,1){8.67}}
\put(119.67,51.00){\vector(-2,1){18.00}}
\put(10.00,5.00){\makebox(0,0)[cc]{$p_1$}}
\put(20.00,5.00){\makebox(0,0)[cc]{$p_2$}}
\put(30.00,5.00){\makebox(0,0)[cc]{$p_3$}}
\put(40.00,5.00){\makebox(0,0)[cc]{$p_4$}}
\put(50.00,5.00){\makebox(0,0)[cc]{$p_5$}}
\put(80.00,5.00){\makebox(0,0)[cc]{$p_1$}}
\put(90.00,5.00){\makebox(0,0)[cc]{$p_2$}}
\put(100.00,5.00){\makebox(0,0)[cc]{$p_3$}}
\put(110.00,5.00){\makebox(0,0)[cc]{$p_4$}}
\put(120.00,5.00){\makebox(0,0)[cc]{$p_5$}}
\put(5.00,95.00){\makebox(0,0)[cc]{a)}}
\put(75.00,95.00){\makebox(0,0)[cc]{b)}}
\put(145.00,95.00){\makebox(0,0)[cc]{c)}}
\put(170.00,90.00){\circle{2.00}}
\put(170.00,80.00){\circle{2.00}}
\put(170.00,70.00){\circle{2.00}}
\put(170.00,60.00){\circle{2.00}}
\put(170.00,50.00){\circle{2.00}}
\put(170.00,40.00){\circle{2.00}}
\put(170.00,30.00){\circle{2.00}}
\put(170.00,20.00){\circle{2.00}}
\put(170.00,10.00){\circle*{2.00}}
\put(170.00,5.00){\makebox(0,0)[cc]{$p_1$}}
\put(170.00,11.00){\vector(0,1){8.00}}
\put(170.00,21.00){\vector(0,1){8.00}}
\put(170.00,31.00){\vector(0,1){8.00}}
\put(170.00,41.00){\vector(0,1){8.00}}
\put(170.00,51.00){\vector(0,1){8.00}}
\put(170.00,61.00){\vector(0,1){8.00}}
\put(170.00,71.00){\vector(0,1){8.00}}
\put(170.00,81.00){\vector(0,1){8.00}}
\put(180.00,70.00){\circle{2.00}}
\put(190.00,70.00){\circle{2.00}}
\put(180.00,80.00){\circle{2.00}}
\put(180.00,71.00){\vector(0,1){8.00}}
\put(180.00,90.00){\circle{2.00}}
\put(180.00,81.00){\vector(0,1){8.00}}
\put(190.00,80.00){\circle{2.00}}
\put(190.00,71.00){\vector(0,1){8.00}}
\put(190.00,90.00){\circle{2.00}}
\put(190.00,81.00){\vector(0,1){8.00}}
\put(150.00,50.00){\circle{2.00}}
\put(150.00,60.00){\circle{2.00}}
\put(150.00,51.00){\vector(0,1){8.00}}
\put(150.00,70.00){\circle{2.00}}
\put(150.00,61.00){\vector(0,1){8.00}}
\put(150.00,80.00){\circle{2.00}}
\put(150.00,71.00){\vector(0,1){8.00}}
\put(150.00,90.00){\circle{2.00}}
\put(150.00,81.00){\vector(0,1){8.00}}
\put(160.00,50.00){\circle{2.00}}
\put(160.00,60.00){\circle{2.00}}
\put(160.00,51.00){\vector(0,1){8.00}}
\put(160.00,70.00){\circle{2.00}}
\put(160.00,61.00){\vector(0,1){8.00}}
\put(160.00,80.00){\circle{2.00}}
\put(160.00,71.00){\vector(0,1){8.00}}
\put(160.00,90.00){\circle{2.00}}
\put(160.00,81.00){\vector(0,1){8.00}}
\put(168.67,39.67){\vector(-2,1){18.00}}
\put(169.00,40.33){\vector(-1,1){8.33}}
\put(170.67,60.33){\vector(1,1){8.67}}
\put(170.67,59.67){\vector(2,1){18.67}}
\end{picture}
\end{center}
\caption{In this flow diagram, the lowest ``root'' represents the
initial state of the computer. Forward computation represents
upwards motion
through a sequence of states represented by open circles. Different
symbols $p_i$ correspond to different initial computer states.
a) One-to-one computation.
b) Many-to-one junction which is information discarding. Several
computational paths, moving upwards, merge into one.
c) One-to-many computation is allowed only
 if no information is
created and discarded; e.g., in copy-type operations on blank memory.
From Landauer \protect\cite{landauer-94}.
\label{f-rev-comp}
}
\end{figure}

The connection between information and physical entropy, in particular
the entropy increase during computational steps corresponding to an
irreversible loss of information---deletion or other many-to-one
operations---has raised considerable attention in the physics community
\cite{maxwell-demon}.
Figure \ref{f-rev-comp}  \cite{landauer-94} depicts a flow diagram,
illustrating the
difference between one-to-one, many-to-one and one-to-many computation.
Classical
reversible computation
\cite{landauer:61,bennett-73,fred-tof-82,bennett-82,landauer-94}
is characterized by a single-valued invertible (i.e., bijective or
one-to-one) evolution function.
In such cases
it is always possible to ``reverse the gear'' of the evolution, and
compute the input from the output, the initial state from the final
state.

In irreversible computations, logical functions
are performed which
do not have a single-valued inverse, such as ${and}$ or ${or}$;
i.e., the input cannot be deduced from the output. Also deletion of
information or other many
(states)-to-one
(state) operations are irreversible.
This logical irreversibility is associated with physical irreversibility
and requires a minimal heat generation of the computing machine and
thus an entropy increase.

It is possible to embed any irreversible computation in an appropriate
environment which makes it reversible. For instance, the computer
could keep the inputs of previous calculations in successive order.
It could save all the information it would otherwise throw away.
 Or,
it could leave markers behind to identify its trail, the {\it H\"ansel
and Gretel} strategy described by Landauer \cite{landauer-94}. That, of
course, might amount to huge overheads in dynamical memory space
(and time) and would merely postpone the problem of throwing away
unwanted information. But, as has been pointed out by Bennett
\cite{bennett-73}, for classical computations, in which copying and
one-to-many operations are still allowed, this overhead could be
circumvented by
erasing all intermediate results, leaving behind only copies of the
output and the original input. Bennett's trick is
to perform  a computation,  copy the resulting output
and then, with one output as input, run
the computation backward. In order not to consume exceedingly large
intermediate storage resources, this strategy could be applied after
every single step.
Notice that copying can be done
reversible in classical physics if the memory used for the copy is
initially considered to be blank.

Quantum mechanics, in particular quantum computing, teaches us to
restrict ourselves even more and exclude any one-to-many operations, in
particular copying, and to accept merely one-to-one
computational operations
corresponding to bijective mappings [cf.
Figure \ref{f-rev-comp}a)].
This is due to the fact that the unitary
evolution of the quantum mechanical state state (between two subsequent
measurements) is strictly one-to-one.
Per definition, the inverse of a unitary operator $U$ representing a
quantum mechanical time evolution always exists. It is again a unitary
operator $U^{-1}=U^\dagger$ (where $\dagger$ represents the adjoint
operator); i.e., $UU^\dagger =1$.
As a consequence, the {\em no-cloning theorem}
\cite{herbert,wo-zu,mandel:83,mil-hard,glauber,caves}
states that certain one-to-many
operations are not allowed, in particular the copying of general
(nonclassical)
quantum bits of information.

In what follows we shall consider a particular example of a
one-to-one deterministic computation.  Although tentative in its present
form, this example may illustrate the conceptual strength of reversible
computation.  Our starting point are
finite automata \cite{e-f-moore,conway,brauer-84,schaller-96,cal-sv-yu},
but of a very particular,
hitherto unknown sort.  They are characterized by a finite set $S$ of
states, a finite input and output alphabet $I$ and $O$, respectively.
Like for Mealy automata, their temporal evolution and output functions
are given by $\delta :S\times I\rightarrow S$, $\lambda :S\times
I\rightarrow O$.  We additionally require one-to-one reversibility,
which we interpret in this context as follows.  Let $I=O$, and let the
combined (state and output) temporal evolution be associated with a
bijective map
\begin{equation}
U:(s,i)\rightarrow (\delta(s,i),\lambda (s,i)),
\label{t-e-l}
\end{equation}
with
$s\in S$ and $i\in I$.
The state and output symbol could be ``fed back'' consecutively; such
that
$N$ evolution steps correspond to $U^N=\underbrace{U\cdots U}_{N
\;\textrm{times}}$.

The elements of the Cartesian product
$S\times I$ can be arranged as a linear list of length
$n$ corresponding to  a vector. In this sense,
$U$ corresponds to a $n\times n$-matrix.  Let $\Psi_i$
be the $i$'th element in the vectorial representation of some
$(s,i)$, and let
$U_{ij}$ be the element
of
$U$ in the $i$'th row and the $j$'th column.
Due to
determinism, uniqueness and invertibility,
\begin{description}
\item[$\bullet$]
$U_{ij}\in \{0,1\}$;
\item[$\bullet$]
orthogonality:
 $U^{-1}=U^t$ (superscript $t$ means transposition) and
$(U^{-1})_{ij}=U_{ji}$;
\item[$\bullet$]
double stochasticity:
the sum of each row and column is one; i.e.,
$\sum_{i=1}^n U_{ij}= \sum_{j=1}^n U_{ij}=1$.
\end{description}
Since $U$ is a square matrix whose elements are either one or zero and
which has exactly one nonzero entry in each row and exactly one in each
column, it is a {\em permutation matrix}.
Let ${\cal P}_n$ denote the set of all $n\times n$ permutation matrices.
 ${\cal P}_n$ forms the {\em permutation group} (sometimes referred to
as the
{\em symmetric group}) of degree $n$. (The product of two permutation
matrices is
a permutation matrix, the inverse is the transpose and the identity
${\bf
1}$ belongs to  ${\cal P}_n$.)
${\cal P}_n$ has $n!$ elements.
Furthermore, the set of all doubly stochastic matrices forms a convex
polyhedron with the permutation matrices as vertices
\cite[page 82]{ber-ple}.

Let us be more specific.
For $n=1$, ${\cal P}_1=\{1\}$.\\
For $n=2$, ${\cal P}_2=\left\{
\left(
\begin{array}{cc}
1&0\\
0&1
\end{array}
\right)
,
\left(
\begin{array}{cc}
0&1\\
1&0
\end{array}
\right)
\right\}$.\\
For $n=3$,
$${\cal P}_3=\left\{
\left(
\begin{array}{ccc}
1&0&0\\
0&1&0\\
0&0&1
\end{array}
\right)
,
\left(
\begin{array}{ccc}
1&0&0\\
0&0&1\\
0&1&0
\end{array}
\right)
,
\left(
\begin{array}{ccc}
0&1&0\\
0&0&1\\
1&0&0
\end{array}
\right)
,\right.$$
$$\qquad \left.
\left(
\begin{array}{ccc}
0&1&0\\
1&0&0\\
0&0&1
\end{array}
\right)
,
\left(
\begin{array}{ccc}
0&0&1\\
1&0&0\\
0&1&0
\end{array}
\right)
,
\left(
\begin{array}{ccc}
0&0&1\\
0&1&0\\
1&0&0
\end{array}
\right)\right\}
.           $$

\begin{figure}
\begin{center}
\unitlength 1.20mm
\linethickness{0.4pt}
\begin{picture}(30.75,15.75)
\put(0.00,5.00){\circle{1.50}}
\put(10.00,5.00){\circle{1.50}}
\put(20.00,5.00){\circle{1.50}}
\put(30.00,5.00){\circle{1.50}}
\put(0.00,15.00){\circle{1.50}}
\put(10.00,15.00){\circle{1.50}}
\put(20.00,15.00){\circle{1.50}}
\put(30.00,15.00){\circle{1.50}}
\put(0.00,0.00){\makebox(0,0)[cc]{$(s_1,1)$}}
\put(10.00,0.00){\makebox(0,0)[cc]{$(s_1,2)$}}
\put(20.00,0.00){\makebox(0,0)[cc]{$(s_2,1)$}}
\put(30.00,0.00){\makebox(0,0)[cc]{$(s_2,2)$}}
\put(19.45,14.45){\vector(4,3){0.2}}
\multiput(0.61,5.48)(0.49,0.11){5}{\line(1,0){0.49}}
\multiput(3.07,6.05)(0.40,0.11){6}{\line(1,0){0.40}}
\multiput(5.45,6.73)(0.33,0.11){7}{\line(1,0){0.33}}
\multiput(7.75,7.52)(0.28,0.11){8}{\line(1,0){0.28}}
\multiput(9.97,8.44)(0.24,0.11){9}{\line(1,0){0.24}}
\multiput(12.10,9.47)(0.21,0.11){10}{\line(1,0){0.21}}
\multiput(14.15,10.62)(0.18,0.11){11}{\line(1,0){0.18}}
\multiput(16.12,11.88)(0.15,0.12){22}{\line(1,0){0.15}}
\put(9.99,13.96){\vector(0,1){0.2}}
\put(9.99,5.81){\line(0,1){8.15}}
\put(0.61,14.37){\vector(-4,3){0.2}}
\multiput(29.24,5.32)(-1.09,0.11){3}{\line(-1,0){1.09}}
\multiput(25.98,5.66)(-0.77,0.10){4}{\line(-1,0){0.77}}
\multiput(22.89,6.07)(-0.73,0.12){4}{\line(-1,0){0.73}}
\multiput(19.96,6.54)(-0.55,0.11){5}{\line(-1,0){0.55}}
\multiput(17.20,7.09)(-0.43,0.10){6}{\line(-1,0){0.43}}
\multiput(14.61,7.70)(-0.40,0.11){6}{\line(-1,0){0.40}}
\multiput(12.18,8.38)(-0.32,0.11){7}{\line(-1,0){0.32}}
\multiput(9.92,9.13)(-0.30,0.12){7}{\line(-1,0){0.30}}
\multiput(7.83,9.95)(-0.24,0.11){8}{\line(-1,0){0.24}}
\multiput(5.91,10.84)(-0.22,0.12){8}{\line(-1,0){0.22}}
\multiput(4.15,11.80)(-0.18,0.11){9}{\line(-1,0){0.18}}
\multiput(2.56,12.82)(-0.15,0.12){13}{\line(-1,0){0.15}}
\put(29.49,14.37){\vector(3,2){0.2}}
\multiput(20.44,5.73)(0.12,0.27){6}{\line(0,1){0.27}}
\multiput(21.16,7.35)(0.11,0.15){11}{\line(0,1){0.15}}
\multiput(22.36,8.99)(0.12,0.12){14}{\line(1,0){0.12}}
\multiput(24.05,10.65)(0.16,0.12){14}{\line(1,0){0.16}}
\multiput(26.23,12.33)(0.19,0.12){17}{\line(1,0){0.19}}
\end{picture}
\end{center}
\caption{Flow diagram of one evolution cycle of the reversible automaton
listed in Table
\protect\ref{t-ra}.
\label{f-fdia}
}
\end{figure}
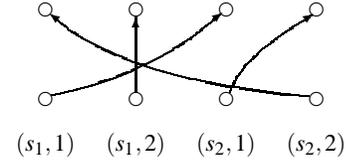

The correspondence between permutation matrices and
reversible automata is straightforward. Per definition [cf. Equation
(\ref{t-e-l})], every reversible automaton is representable by some
permutation matrix. That every $n\times n$ permutation matrix
corresponds to an automaton can be demonstrated by
considering the simplest
case of a one state automaton with $n$ input/output symbols.
There exist less trivial identifications. For example,
let $$  {U}=
\left(
\begin{array}{cccc}
0&0&1&0\\
0&1&0&0\\
0&0&0&1\\
1&0&0&0
\end{array}
\right).
$$
The transition and output functions of one associated reversible
automaton is listed in table
\ref{t-ra}.
\begin{table}
\begin{center}
\begin{tabular}{|c|cc|cc|}
 \hline
 \hline
 &$\delta$ & & $\lambda$&\\
$S\backslash I$ &1&2& 1&2\\
 \hline
$s_1$&$s_2$&$s_1$ & 1&2\\
$s_2$&$s_2$&$s_1 $& 2&1\\
 \hline
 \hline
\end{tabular}
\end{center}
\caption{Transition and output table of a reversible
automaton with two states $S=\{s_1, s_2\}$ and two input/output
symbols $I= \{1,2\}$.\label{t-ra}}
\end{table}
The associated flow diagram is drawn in Figure \ref{f-fdia}.
Since ${U}$ has a cycle 3; i.e., $({U})^3={\bf 1}$,
irrespective of the initial state, the automaton is back at its initial
state after three evolution steps. For example,
$
(s_1,1)\rightarrow
(s_2,1)\rightarrow
(s_2,2)\rightarrow
(s_1,1)$.

The discrete temporal evolution (\ref{t-e-l}) can, in matrix
notation, be represented by
\begin{equation}
U\Psi (N)= \Psi (N+1)=U^{N+1}\Psi (0),
\label{t-e-l2}
\end{equation}
where again $N=0,1,2,3,\ldots$ is a discrete time parameter.



\begin{figure}
\begin{center}
\unitlength 0.9mm
\linethickness{0.4pt}
\begin{picture}(79.33,115.00)
\put(2.33,14.67){\framebox(5.00,5.00)[cc]{$\footnotesize i_{1}$}}
\put(9.33,14.67){\framebox(5.00,5.00)[cc]{$\footnotesize i_{2}$}}
\put(16.33,14.67){\framebox(5.00,5.00)[cc]{$\footnotesize i_{3}$}}
\put(23.33,14.67){\framebox(5.00,5.00)[cc]{$\footnotesize i_{4}$}}
\put(30.33,14.67){\framebox(5.00,5.00)[cc]{$\footnotesize i_{5}$}}
\put(37.33,14.67){\framebox(5.00,5.00)[cc]{$\footnotesize i_{6}$}}
\put(44.33,14.67){\framebox(5.00,5.00)[cc]{$\footnotesize i_{7}$}}
\put(51.33,14.67){\framebox(5.00,5.00)[cc]{$\footnotesize i_{8}$}}
\put(58.33,14.67){\framebox(5.00,5.00)[cc]{$\footnotesize i_{9}$}}
\put(65.33,14.67){\framebox(5.00,5.00)[cc]{$\footnotesize i_{10}$}}
\put(72.33,14.67){\framebox(5.00,5.00)[cc]{$\footnotesize i_{11}$}}
\put(2.33,7.67){\framebox(5.00,5.00)[cc]{$\footnotesize i_{12}$}}
\put(9.33,7.67){\framebox(5.00,5.00)[cc]{$\footnotesize i_{13}$}}
\put(16.33,7.67){\framebox(5.00,5.00)[cc]{$\footnotesize i_{14}$}}
\put(23.33,7.67){\framebox(5.00,5.00)[cc]{$\footnotesize i_{15}$}}
\put(30.33,7.67){\framebox(5.00,5.00)[cc]{$\footnotesize i_{16}$}}
\put(37.33,7.67){\framebox(5.00,5.00)[cc]{$\footnotesize i_{17}$}}
\put(44.33,7.67){\framebox(5.00,5.00)[cc]{$\footnotesize i_{18}$}}
\put(51.33,7.67){\framebox(5.00,5.00)[cc]{$\footnotesize i_{19}$}}
\put(58.33,7.67){\framebox(5.00,5.00)[cc]{$\cdot$}}
\put(65.33,7.67){\framebox(5.00,5.00)[cc]{$\cdot$}}
\put(72.33,7.67){\framebox(5.00,5.00)[cc]{$\cdot$}}
\put(20.00,35.00){\line(1,0){30.00}}
\put(50.00,35.00){\line(0,1){25.00}}
\put(50.00,60.00){\line(-1,0){30.00}}
\put(20.00,60.00){\line(0,-1){25.00}}
\put(50.00,60.00){\line(3,2){15.00}}
\put(20.00,60.00){\line(3,2){15.00}}
\put(50.00,35.00){\line(3,2){15.00}}
\put(65.00,45.00){\line(0,1){25.00}}
\put(65.00,70.00){\line(-1,0){30.00}}
\put(0.00,5.00){\line(1,0){79.33}}
\put(79.33,5.00){\line(0,1){17.00}}
\put(79.33,22.00){\line(-1,0){79.33}}
\put(0.00,22.00){\line(0,-1){17.00}}
\put(30.00,90.00){\line(1,0){20.00}}
\put(50.00,90.00){\line(0,1){20.00}}
\put(50.00,110.00){\line(-1,0){20.00}}
\put(30.00,110.00){\line(0,-1){20.00}}
\multiput(40.67,66.33)(0.11,0.16){15}{\line(0,1){0.16}}
\multiput(42.35,68.80)(0.12,0.20){11}{\line(0,1){0.20}}
\multiput(43.67,70.99)(0.12,0.24){8}{\line(0,1){0.24}}
\multiput(44.62,72.90)(0.12,0.33){5}{\line(0,1){0.33}}
\multiput(45.20,74.55)(0.11,0.69){2}{\line(0,1){0.69}}
\multiput(45.42,75.92)(-0.08,0.55){2}{\line(0,1){0.55}}
\multiput(45.27,77.01)(-0.10,0.16){5}{\line(0,1){0.16}}
\multiput(44.75,77.84)(-0.18,0.11){5}{\line(-1,0){0.18}}
\multiput(43.87,78.39)(-0.42,0.09){3}{\line(-1,0){0.42}}
\put(42.62,78.66){\line(-1,0){1.62}}
\put(41.00,78.67){\line(-1,0){1.85}}
\multiput(39.15,78.71)(-0.50,0.08){3}{\line(-1,0){0.50}}
\multiput(37.66,78.95)(-0.28,0.11){4}{\line(-1,0){0.28}}
\multiput(36.53,79.41)(-0.13,0.11){6}{\line(-1,0){0.13}}
\multiput(35.78,80.07)(-0.10,0.22){4}{\line(0,1){0.22}}
\put(35.39,80.93){\line(0,1){1.07}}
\multiput(35.36,82.00)(0.11,0.43){3}{\line(0,1){0.43}}
\multiput(35.70,83.28)(0.12,0.25){6}{\line(0,1){0.25}}
\multiput(36.41,84.76)(0.12,0.19){9}{\line(0,1){0.19}}
\multiput(37.48,86.44)(0.12,0.15){24}{\line(0,1){0.15}}
\multiput(39.67,34.67)(0.18,-0.11){9}{\line(1,0){0.18}}
\multiput(41.31,33.67)(0.11,-0.13){8}{\line(0,-1){0.13}}
\put(42.18,32.59){\line(0,-1){1.16}}
\multiput(42.29,31.43)(-0.11,-0.21){6}{\line(0,-1){0.21}}
\multiput(41.64,30.18)(-0.14,-0.11){19}{\line(-1,0){0.14}}
\multiput(39.00,28.00)(-0.22,-0.12){8}{\line(-1,0){0.22}}
\multiput(37.23,27.05)(-0.12,-0.13){8}{\line(0,-1){0.13}}
\multiput(36.27,26.01)(-0.07,-0.57){2}{\line(0,-1){0.57}}
\multiput(36.12,24.86)(0.11,-0.21){6}{\line(0,-1){0.21}}
\multiput(36.79,23.62)(0.13,-0.12){14}{\line(1,0){0.13}}
\put(35.00,47.33){\makebox(0,0)[cc]{\huge ?}}
\put(40.00,100.00){\makebox(0,0)[cc]{$o_j$}}
\put(40.00,0.00){\makebox(0,0)[cc]{input interface}}
\put(40.00,115.00){\makebox(0,0)[cc]{output interface}}
\put(70.00,55.00){\makebox(0,0)[lc]{box}}
\end{picture}
\end{center}
\caption{\label{f-bbox}
A system in a black box
with an input interface and an output interface.}
\end{figure}
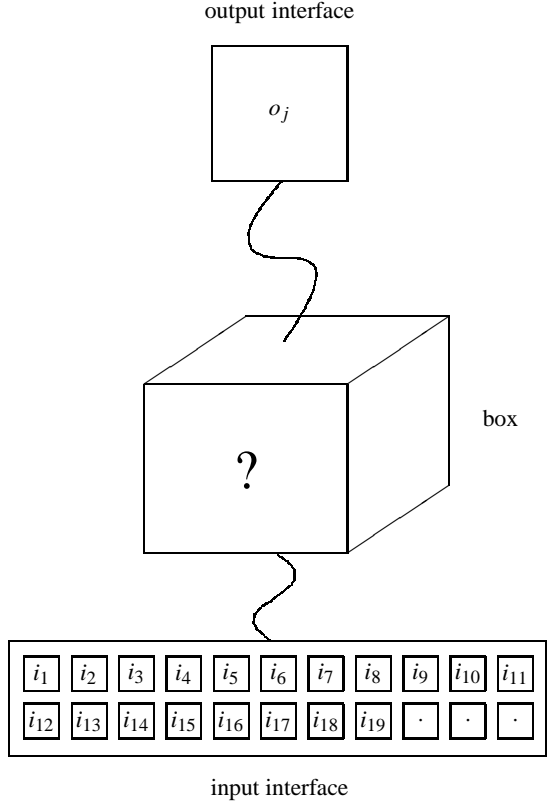

Let us come back to our original issue of modelling the measurement
process within a system whose states evolve according  to a one-to-one
evolution.
Let us artificially divide such a system into an ``inside'' and an
``outside'' region.
This can be suitably represented by introducing a black box which
contains the ``inside'' region---the subsystem to be measured, whereas
the
remaining ``outside'' region is interpreted as the measurement
apparatus.
An input and an output interface mediate all interactions of the
``inside'' with the ``outside,'' of the ``observed'' and the
``observer'' by symbolic exchange. Let us assume that, despite such
symbolic exchanges via the interfaces
(for all practical purposes),
to an outside
observer
 what happens inside the black box is a hidden,
 inaccessible
  arena.
The observed system is like the ``black box'' drawn in
Figure~\ref{f-bbox}.

Throughout temporal evolution, not only is information transformed
one-to-one
(bijectively, homomorphically) inside the black box,
but
this information is handled  one-to-one {\em after} it appeared on the
black box
interfaces. It might  seem evident at first glance that the symbols
appearing on the interfaces should be treated as classical information.
That is, they could in principle be copied. The possibility to copy the
experiment (input and output) enables the application of Bennett's
argument: in such a case, one keeps the experimental finding by copying
it, reverts the system evolution and starts with a ``fresh'' black
box system in its original initial state. The result is a
classical Boolean calculus.

The scenario is drastically changed, however, if we assume a
one-to-one evolution also for the environment at and outside
of the black box. That is, one deals with a
homogeneous
and uniform one-to-one evolution ``inside'' and ``outside'' of the black
box, thereby
assuming that the experimenter also evolves
one-to-one and not classically.
In our toy automaton model, this could for instance be realized by some
automaton corresponding to a permutation operator $U$ inside the black
box, and another reversible automaton corresponding to another $U'$
outside of it. Conventionally, $U$ and $U'$ correspond to the measured
system and the measurement device, respectively.

In such a case, as there is no copying due to one-to-one evolution,
in order to set
back the system to its original initial state, the experimenter would
have to erase all knowledge bits of information acquired so far.
The experiment would have to evolve back to the initial state of the
measurement device and the measured system prior to the measurement.
As a result, the representation of measurement results in one-to-one
reversible systems may cause a sort of complementarity due to
the impossibility to measure all variants of the representation
at once.

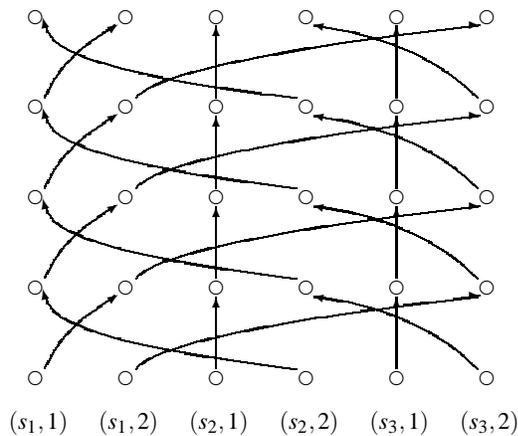
\begin{figure}
\begin{center}
\unitlength 1.20mm
\linethickness{0.4pt}
\begin{picture}(50.75,45.80)
\put(0.00,5.00){\circle{1.50}}
\put(10.00,5.00){\circle{1.50}}
\put(20.00,5.00){\circle{1.50}}
\put(30.00,5.00){\circle{1.50}}
\put(0.00,15.00){\circle{1.50}}
\put(10.00,15.00){\circle{1.50}}
\put(20.00,15.00){\circle{1.50}}
\put(30.00,15.00){\circle{1.50}}
\put(0.00,0.00){\makebox(0,0)[cc]{ $(s_1,1)$}}
\put(10.00,0.00){\makebox(0,0)[cc]{ $(s_1,2)$}}
\put(20.00,0.00){\makebox(0,0)[cc]{ $(s_2,1)$}}
\put(30.00,0.00){\makebox(0,0)[cc]{ $(s_2,2)$}}
\put(40.00,5.00){\circle{1.50}}
\put(50.00,5.00){\circle{1.50}}
\put(40.00,15.00){\circle{1.50}}
\put(50.00,15.00){\circle{1.50}}
\put(40.00,0.00){\makebox(0,0)[cc]{ $(s_3,1)$}}
\put(50.00,0.00){\makebox(0,0)[cc]{ $(s_3,2)$}}
\put(20.00,6.00){\vector(0,1){8.00}}
\put(40.00,6.00){\vector(0,1){8.00}}
\put(8.96,14.09){\vector(3,2){0.2}}
\multiput(1.05,6.00)(0.12,0.23){9}{\line(0,1){0.23}}
\multiput(2.09,8.03)(0.11,0.15){12}{\line(0,1){0.15}}
\multiput(3.45,9.88)(0.12,0.12){14}{\line(1,0){0.12}}
\multiput(5.13,11.55)(0.17,0.12){22}{\line(1,0){0.17}}
\put(49.03,14.09){\vector(1,0){0.2}}
\multiput(11.07,6.00)(0.10,0.10){5}{\line(1,0){0.10}}
\multiput(11.59,6.51)(0.15,0.10){5}{\line(1,0){0.15}}
\multiput(12.35,7.01)(0.20,0.10){5}{\line(1,0){0.20}}
\multiput(13.37,7.52)(0.25,0.10){5}{\line(1,0){0.25}}
\multiput(14.63,8.02)(0.30,0.10){5}{\line(1,0){0.30}}
\multiput(16.14,8.53)(0.35,0.10){5}{\line(1,0){0.35}}
\multiput(17.89,9.03)(0.40,0.10){5}{\line(1,0){0.40}}
\multiput(19.89,9.54)(0.45,0.10){5}{\line(1,0){0.45}}
\multiput(22.14,10.04)(0.50,0.10){5}{\line(1,0){0.50}}
\multiput(24.64,10.55)(0.55,0.10){5}{\line(1,0){0.55}}
\multiput(27.38,11.06)(0.60,0.10){5}{\line(1,0){0.60}}
\multiput(30.37,11.56)(0.65,0.10){5}{\line(1,0){0.65}}
\multiput(33.61,12.07)(0.80,0.11){9}{\line(1,0){0.80}}
\multiput(40.82,13.08)(0.91,0.11){9}{\line(1,0){0.91}}
\put(1.05,14.09){\vector(-1,4){0.2}}
\multiput(28.99,6.00)(-0.81,0.10){5}{\line(-1,0){0.81}}
\multiput(24.95,6.52)(-0.75,0.11){5}{\line(-1,0){0.75}}
\multiput(21.21,7.05)(-0.68,0.11){5}{\line(-1,0){0.68}}
\multiput(17.80,7.59)(-0.62,0.11){5}{\line(-1,0){0.62}}
\multiput(14.70,8.16)(-0.56,0.12){5}{\line(-1,0){0.56}}
\multiput(11.92,8.73)(-0.49,0.12){5}{\line(-1,0){0.49}}
\multiput(9.45,9.33)(-0.36,0.10){6}{\line(-1,0){0.36}}
\multiput(7.30,9.94)(-0.31,0.10){6}{\line(-1,0){0.31}}
\multiput(5.47,10.57)(-0.25,0.11){6}{\line(-1,0){0.25}}
\multiput(3.96,11.21)(-0.20,0.11){6}{\line(-1,0){0.20}}
\multiput(2.76,11.87)(-0.15,0.11){6}{\line(-1,0){0.15}}
\multiput(1.88,12.54)(-0.11,0.14){5}{\line(0,1){0.14}}
\multiput(1.31,13.24)(-0.09,0.28){3}{\line(0,1){0.28}}
\put(30.93,14.09){\vector(-4,1){0.2}}
\multiput(49.03,6.00)(-0.12,0.11){13}{\line(-1,0){0.12}}
\multiput(47.44,7.48)(-0.14,0.11){12}{\line(-1,0){0.14}}
\multiput(45.71,8.82)(-0.19,0.12){10}{\line(-1,0){0.19}}
\multiput(43.81,10.02)(-0.23,0.12){9}{\line(-1,0){0.23}}
\multiput(41.77,11.08)(-0.28,0.12){8}{\line(-1,0){0.28}}
\multiput(39.57,12.00)(-0.34,0.11){7}{\line(-1,0){0.34}}
\multiput(37.21,12.78)(-0.42,0.11){6}{\line(-1,0){0.42}}
\multiput(34.70,13.42)(-0.63,0.11){6}{\line(-1,0){0.63}}
\put(0.00,25.02){\circle{1.50}}
\put(0.00,35.03){\circle{1.50}}
\put(0.00,45.05){\circle{1.50}}
\put(10.00,25.02){\circle{1.50}}
\put(10.00,35.03){\circle{1.50}}
\put(10.00,45.05){\circle{1.50}}
\put(20.00,25.02){\circle{1.50}}
\put(20.00,35.03){\circle{1.50}}
\put(20.00,45.05){\circle{1.50}}
\put(30.00,25.02){\circle{1.50}}
\put(30.00,35.03){\circle{1.50}}
\put(30.00,45.05){\circle{1.50}}
\put(40.00,25.02){\circle{1.50}}
\put(40.00,35.03){\circle{1.50}}
\put(40.00,45.05){\circle{1.50}}
\put(50.00,25.02){\circle{1.50}}
\put(50.00,35.03){\circle{1.50}}
\put(50.00,45.05){\circle{1.50}}
\put(20.00,16.02){\vector(0,1){8.00}}
\put(20.00,26.03){\vector(0,1){8.00}}
\put(20.00,36.05){\vector(0,1){8.00}}
\put(40.00,16.02){\vector(0,1){8.00}}
\put(40.00,26.03){\vector(0,1){8.00}}
\put(40.00,36.05){\vector(0,1){8.00}}
\put(8.96,24.10){\vector(3,2){0.2}}
\multiput(1.05,16.02)(0.12,0.23){9}{\line(0,1){0.23}}
\multiput(2.09,18.05)(0.11,0.15){12}{\line(0,1){0.15}}
\multiput(3.45,19.90)(0.12,0.12){14}{\line(1,0){0.12}}
\multiput(5.13,21.57)(0.17,0.12){22}{\line(1,0){0.17}}
\put(8.96,34.12){\vector(3,2){0.2}}
\multiput(1.05,26.04)(0.12,0.23){9}{\line(0,1){0.23}}
\multiput(2.09,28.07)(0.11,0.15){12}{\line(0,1){0.15}}
\multiput(3.45,29.92)(0.12,0.12){14}{\line(1,0){0.12}}
\multiput(5.13,31.58)(0.17,0.12){22}{\line(1,0){0.17}}
\put(8.96,44.14){\vector(3,2){0.2}}
\multiput(1.05,36.05)(0.12,0.23){9}{\line(0,1){0.23}}
\multiput(2.09,38.08)(0.11,0.15){12}{\line(0,1){0.15}}
\multiput(3.45,39.93)(0.12,0.12){14}{\line(1,0){0.12}}
\multiput(5.13,41.60)(0.17,0.12){22}{\line(1,0){0.17}}
\put(49.03,24.10){\vector(1,0){0.2}}
\multiput(11.07,16.02)(0.10,0.10){5}{\line(1,0){0.10}}
\multiput(11.59,16.52)(0.15,0.10){5}{\line(1,0){0.15}}
\multiput(12.35,17.03)(0.20,0.10){5}{\line(1,0){0.20}}
\multiput(13.37,17.54)(0.25,0.10){5}{\line(1,0){0.25}}
\multiput(14.63,18.04)(0.30,0.10){5}{\line(1,0){0.30}}
\multiput(16.14,18.55)(0.35,0.10){5}{\line(1,0){0.35}}
\multiput(17.89,19.05)(0.40,0.10){5}{\line(1,0){0.40}}
\multiput(19.89,19.56)(0.45,0.10){5}{\line(1,0){0.45}}
\multiput(22.14,20.06)(0.50,0.10){5}{\line(1,0){0.50}}
\multiput(24.64,20.57)(0.55,0.10){5}{\line(1,0){0.55}}
\multiput(27.38,21.07)(0.60,0.10){5}{\line(1,0){0.60}}
\multiput(30.37,21.58)(0.65,0.10){5}{\line(1,0){0.65}}
\multiput(33.61,22.08)(0.80,0.11){9}{\line(1,0){0.80}}
\multiput(40.82,23.09)(0.91,0.11){9}{\line(1,0){0.91}}
\put(49.03,34.12){\vector(1,0){0.2}}
\multiput(11.07,26.04)(0.10,0.10){5}{\line(1,0){0.10}}
\multiput(11.59,26.54)(0.15,0.10){5}{\line(1,0){0.15}}
\multiput(12.35,27.05)(0.20,0.10){5}{\line(1,0){0.20}}
\multiput(13.37,27.55)(0.25,0.10){5}{\line(1,0){0.25}}
\multiput(14.63,28.06)(0.30,0.10){5}{\line(1,0){0.30}}
\multiput(16.14,28.56)(0.35,0.10){5}{\line(1,0){0.35}}
\multiput(17.89,29.07)(0.40,0.10){5}{\line(1,0){0.40}}
\multiput(19.89,29.57)(0.45,0.10){5}{\line(1,0){0.45}}
\multiput(22.14,30.08)(0.50,0.10){5}{\line(1,0){0.50}}
\multiput(24.64,30.58)(0.55,0.10){5}{\line(1,0){0.55}}
\multiput(27.38,31.09)(0.60,0.10){5}{\line(1,0){0.60}}
\multiput(30.37,31.59)(0.65,0.10){5}{\line(1,0){0.65}}
\multiput(33.61,32.10)(0.80,0.11){9}{\line(1,0){0.80}}
\multiput(40.82,33.11)(0.91,0.11){9}{\line(1,0){0.91}}
\put(49.03,44.14){\vector(1,0){0.2}}
\multiput(11.07,36.05)(0.10,0.10){5}{\line(1,0){0.10}}
\multiput(11.59,36.56)(0.15,0.10){5}{\line(1,0){0.15}}
\multiput(12.35,37.06)(0.20,0.10){5}{\line(1,0){0.20}}
\multiput(13.37,37.57)(0.25,0.10){5}{\line(1,0){0.25}}
\multiput(14.63,38.07)(0.30,0.10){5}{\line(1,0){0.30}}
\multiput(16.14,38.58)(0.35,0.10){5}{\line(1,0){0.35}}
\multiput(17.89,39.08)(0.40,0.10){5}{\line(1,0){0.40}}
\multiput(19.89,39.59)(0.45,0.10){5}{\line(1,0){0.45}}
\multiput(22.14,40.09)(0.50,0.10){5}{\line(1,0){0.50}}
\multiput(24.64,40.60)(0.55,0.10){5}{\line(1,0){0.55}}
\multiput(27.38,41.11)(0.60,0.10){5}{\line(1,0){0.60}}
\multiput(30.37,41.61)(0.65,0.10){5}{\line(1,0){0.65}}
\multiput(33.61,42.12)(0.80,0.11){9}{\line(1,0){0.80}}
\multiput(40.82,43.13)(0.91,0.11){9}{\line(1,0){0.91}}
\put(1.05,24.10){\vector(-1,4){0.2}}
\multiput(28.99,16.02)(-0.81,0.10){5}{\line(-1,0){0.81}}
\multiput(24.95,16.53)(-0.75,0.11){5}{\line(-1,0){0.75}}
\multiput(21.21,17.06)(-0.68,0.11){5}{\line(-1,0){0.68}}
\multiput(17.80,17.61)(-0.62,0.11){5}{\line(-1,0){0.62}}
\multiput(14.70,18.17)(-0.56,0.12){5}{\line(-1,0){0.56}}
\multiput(11.92,18.75)(-0.49,0.12){5}{\line(-1,0){0.49}}
\multiput(9.45,19.35)(-0.36,0.10){6}{\line(-1,0){0.36}}
\multiput(7.30,19.96)(-0.31,0.10){6}{\line(-1,0){0.31}}
\multiput(5.47,20.58)(-0.25,0.11){6}{\line(-1,0){0.25}}
\multiput(3.96,21.23)(-0.20,0.11){6}{\line(-1,0){0.20}}
\multiput(2.76,21.89)(-0.15,0.11){6}{\line(-1,0){0.15}}
\multiput(1.88,22.56)(-0.11,0.14){5}{\line(0,1){0.14}}
\multiput(1.31,23.25)(-0.09,0.28){3}{\line(0,1){0.28}}
\put(1.05,34.12){\vector(-1,4){0.2}}
\multiput(28.99,26.04)(-0.81,0.10){5}{\line(-1,0){0.81}}
\multiput(24.95,26.55)(-0.75,0.11){5}{\line(-1,0){0.75}}
\multiput(21.21,27.08)(-0.68,0.11){5}{\line(-1,0){0.68}}
\multiput(17.80,27.63)(-0.62,0.11){5}{\line(-1,0){0.62}}
\multiput(14.70,28.19)(-0.56,0.12){5}{\line(-1,0){0.56}}
\multiput(11.92,28.77)(-0.49,0.12){5}{\line(-1,0){0.49}}
\multiput(9.45,29.36)(-0.36,0.10){6}{\line(-1,0){0.36}}
\multiput(7.30,29.97)(-0.31,0.10){6}{\line(-1,0){0.31}}
\multiput(5.47,30.60)(-0.25,0.11){6}{\line(-1,0){0.25}}
\multiput(3.96,31.24)(-0.20,0.11){6}{\line(-1,0){0.20}}
\multiput(2.76,31.90)(-0.15,0.11){6}{\line(-1,0){0.15}}
\multiput(1.88,32.58)(-0.11,0.14){5}{\line(0,1){0.14}}
\multiput(1.31,33.27)(-0.09,0.28){3}{\line(0,1){0.28}}
\put(1.05,44.14){\vector(-1,4){0.2}}
\multiput(28.99,36.05)(-0.81,0.10){5}{\line(-1,0){0.81}}
\multiput(24.95,36.57)(-0.75,0.11){5}{\line(-1,0){0.75}}
\multiput(21.21,37.10)(-0.68,0.11){5}{\line(-1,0){0.68}}
\multiput(17.80,37.64)(-0.62,0.11){5}{\line(-1,0){0.62}}
\multiput(14.70,38.21)(-0.56,0.12){5}{\line(-1,0){0.56}}
\multiput(11.92,38.78)(-0.49,0.12){5}{\line(-1,0){0.49}}
\multiput(9.45,39.38)(-0.36,0.10){6}{\line(-1,0){0.36}}
\multiput(7.30,39.99)(-0.31,0.10){6}{\line(-1,0){0.31}}
\multiput(5.47,40.62)(-0.25,0.11){6}{\line(-1,0){0.25}}
\multiput(3.96,41.26)(-0.20,0.11){6}{\line(-1,0){0.20}}
\multiput(2.76,41.92)(-0.15,0.11){6}{\line(-1,0){0.15}}
\multiput(1.88,42.59)(-0.11,0.14){5}{\line(0,1){0.14}}
\multiput(1.31,43.29)(-0.09,0.28){3}{\line(0,1){0.28}}
\put(30.93,24.10){\vector(-4,1){0.2}}
\multiput(49.03,16.02)(-0.12,0.11){13}{\line(-1,0){0.12}}
\multiput(47.44,17.50)(-0.14,0.11){12}{\line(-1,0){0.14}}
\multiput(45.71,18.84)(-0.19,0.12){10}{\line(-1,0){0.19}}
\multiput(43.81,20.04)(-0.23,0.12){9}{\line(-1,0){0.23}}
\multiput(41.77,21.10)(-0.28,0.12){8}{\line(-1,0){0.28}}
\multiput(39.57,22.02)(-0.34,0.11){7}{\line(-1,0){0.34}}
\multiput(37.21,22.80)(-0.42,0.11){6}{\line(-1,0){0.42}}
\multiput(34.70,23.44)(-0.63,0.11){6}{\line(-1,0){0.63}}
\put(30.93,34.12){\vector(-4,1){0.2}}
\multiput(49.03,26.04)(-0.12,0.11){13}{\line(-1,0){0.12}}
\multiput(47.44,27.51)(-0.14,0.11){12}{\line(-1,0){0.14}}
\multiput(45.71,28.85)(-0.19,0.12){10}{\line(-1,0){0.19}}
\multiput(43.81,30.05)(-0.23,0.12){9}{\line(-1,0){0.23}}
\multiput(41.77,31.11)(-0.28,0.12){8}{\line(-1,0){0.28}}
\multiput(39.57,32.03)(-0.34,0.11){7}{\line(-1,0){0.34}}
\multiput(37.21,32.81)(-0.42,0.11){6}{\line(-1,0){0.42}}
\multiput(34.70,33.46)(-0.63,0.11){6}{\line(-1,0){0.63}}
\put(30.93,44.14){\vector(-4,1){0.2}}
\multiput(49.03,36.05)(-0.12,0.11){13}{\line(-1,0){0.12}}
\multiput(47.44,37.53)(-0.14,0.11){12}{\line(-1,0){0.14}}
\multiput(45.71,38.87)(-0.19,0.12){10}{\line(-1,0){0.19}}
\multiput(43.81,40.07)(-0.23,0.12){9}{\line(-1,0){0.23}}
\multiput(41.77,41.13)(-0.28,0.12){8}{\line(-1,0){0.28}}
\multiput(39.57,42.05)(-0.34,0.11){7}{\line(-1,0){0.34}}
\multiput(37.21,42.83)(-0.42,0.11){6}{\line(-1,0){0.42}}
\multiput(34.70,43.47)(-0.63,0.11){6}{\line(-1,0){0.63}}
\end{picture}
\end{center}
\caption{Flow diagram of four evolution cycles of the reversible
automaton listed in Table
\protect\ref{t-rra}.
\label{f-ffdia}
}
\end{figure}

Let us give a brief example. Consider the $6\times 6$ permutation matrix
$$
U=
\left(
\begin{array}{cccccc}
0&1&0&0&0&0\\
0&0&0&0&0&1\\
0&0&1&0&0&0\\
1&0&0&0&0&0\\
0&0&0&0&1&0\\
0&0&0&1&0&0
\end{array}
\right)
$$
corresponding to a reversible $3$-state automaton with two input/output
symbols $1,2$
listed in table
\ref{t-rra}.
\begin{table}
\begin{center}
\begin{tabular}{|c|cc|cc|}
 \hline
 \hline
 &$\delta$ & & $\lambda$&\\
$S\backslash I$ &1&2& 1&2\\
 \hline
$s_1$&$s_1$&$s_3$ & 2&2\\
$s_2$&$s_2$&$s_1 $& 1&1\\
$s_3$&$s_3$&$s_2 $& 1&2\\
 \hline
 \hline
\end{tabular}
\end{center}
\caption{Transition and output table of a reversible
automaton with three states $S=\{s_1, s_2, s_3\}$ and two input/output
symbols $I= \{1,2\}$.\label{t-rra}}
\end{table}
The evolution is
$$
\left(
\begin{array}{c}
(s_1,1)                \\
(s_1,2)                \\
(s_2,1)                \\
(s_2,2)                \\
(s_3,1)                \\
(s_3,2)
\end{array}
\right)
\stackrel{U}{\longrightarrow }
\left(
\begin{array}{c}
(s_1,2)                \\
(s_3,2)                \\
(s_2,1)                \\
(s_1,1)                \\
(s_3,1)                \\
(s_2,2)
\end{array}
\right)
\stackrel{U}{\longrightarrow }
\left(
\begin{array}{c}
(s_3,2)                \\
(s_2,2)                \\
(s_2,1)                \\
(s_1,2)                \\
(s_3,1)                \\
(s_1,1)
\end{array}
\right)
$$
$$\qquad
\stackrel{U}{\longrightarrow }
\left(
\begin{array}{c}
(s_2,2)                \\
(s_1,1)                \\
(s_2,1)                \\
(s_3,2)                \\
(s_3,1)                \\
(s_1,2)
\end{array}
\right)
\stackrel{U}{\longrightarrow }
\left(
\begin{array}{c}
(s_1,1)                \\
(s_1,2)                \\
(s_2,1)                \\
(s_2,2)                \\
(s_3,1)                \\
(s_3,2)
\end{array}
\right).
  $$
The associated flow diagram is drawn in Figure \ref{f-ffdia}.
Thus after the input of just one symbol, the automaton states can be
grouped into experimental equivalence classes \cite{svozil-93}
$$v(1)=\{\{1\},\{2,3\}\},\quad
v(2)=\{\{1,3\},\{2\}\}.$$
The associated partition logic corresponds to a non Boolean
(nondistributive)
partition logic isomorphic to $MO_2$. Of course, if one develops the
automaton further, then, for instance, $v(2222)=\{\{1\},\{2\},\{3\}\}$,
and the classical case is recovered [notice that this is not the case
for $v(\stackrel{\cdot}{1})=v(1)$]. Yet, if one assumes that the output
is channelled away
into the interface after only a single evolution step (and
that afterwards the evolution is via another $U'$), the
nonclassical feature pertains despite the bijective character of the
evolution.

In this epistemic model, the interface symbolizes the {\em cut} between
the observer and the observed. The cut appears somewhat arbitrary in a
computational universe which is assumed to be uniformly reversible.

What has been discussed above is very similar to the
opening,
closing and reopening of Schr\"odinger's catalogue of expectation values
\cite[p. 53]{schrodinger}:
At least up to
a certain magnitude of complexity---any measurement can be ``undone'' by
a proper reconstruction of the wave-function. A necessary condition for
this to happen is that {\em all} information about the original
measurement is lost.
In Schr\"odinger's terms,
the prediction catalog
(the wave function) can be opened only at one particular page.
We may close
the prediction catalog
before reading this page. Then we can open
the prediction catalog
at another, complementary, page again.
By no way we can open
the prediction catalog at one page, read and (irreversible) memorize the
page, close it; then open it at another, complementary, page.
(Two noncomplementary pages which correspond to two co-measurable
observables can be read simultaneously.)

From this point of view, it appears that, strictly speaking,
irreversibility
may turn out to be an inappropriate concept both in computational
universes generated by one-to-one evolution as well as for quantum
measurement theory. Indeed, irreversibility may have been imposed upon
the measurement process
rather heuristically and artificially to express the huge practical
difficulties associated with any
backward evolution, with ``reversing the gear'', or with reconstructing
a coherent state.
 To quote Landauer
\cite[section 2]{landauer-89},
 \begin{quote}
{\em ``What is measurement? If it is simply information transfer, that
is done all the time inside the computer, and can be done with arbitrary
little dissipation.''}
 \end{quote}

Let us conclude with a metaphysical speculation.
In a one-to-one invertible universe, any  evolution, any step of
computation, any single measurement
act reminds us of a permanent permutation, reformulation and reiteration
of one
and the same ``message''---a ``message'' that was there already at the
beginning of the universe, which gets transformed but is neither
destroyed nor renewed. This thought might be very close to what
Schr\"odinger had in mind when contemplating about Vedic philosophy
\cite{schroed:welt}.


\end{document}